\documentclass[12pt,draftcls,onecolumn]{IEEEtran}

\IEEEoverridecommandlockouts                              

\usepackage{graphics} 
\usepackage{amssymb,amsfonts,amsmath}
\usepackage{epsfig,mathrsfs} 
\usepackage{accents} 
\usepackage{epstopdf}
\usepackage{caption}
\usepackage[usenames]{color}
\newtheorem{lem}{Lemma}
\newtheorem{df}{Definition}
\newtheorem{theo}{Theorem}
\newtheorem{rem}{Remark}

\newcommand{\carre} {\hfill $\blacksquare$}
\newcommand{\carrew} {\hfill $\Box$}

\title{\LARGE \bf
A Predictor-based Attitude and Position Estimation for Rigid Bodies Moving in Planar Space by using Delayed Landmark Measurements
}

\author{Danial Senejohnny  and Mehrzad Namvar
\thanks{}
\thanks{Danial Senejohnny and Mehrzad Namvar are with the Department of Electrical Engineering, Sharif University of Technology, P.O. Box 11155-8639, Tehran, Iran. {\tt\small d.senejohnny@gmail.com, namvar@sharif.ir}}%
}

\begin{document}

\maketitle
\thispagestyle{empty}
\pagestyle{empty}

\begin{abstract}

This paper proposes a globally and exponentially convergent predictive observer for attitude and position estimation based on landmark measurements and velocity (angular and linear) readings. It is assumed that landmark measurements are available with time-delay. The maximum value of the sensor delay under which the estimation error converges to zero is calculated. Synthesis of the observer is based on a representation of rigid-body kinematics and sensor delay, formulated via ordinary and partial differential equations (ODE-PDE). Observability condition specifies necessary and sufficient landmark configuration for convergence of attitude and position estimation error to zero.  Finally, for implementation purposes, a PDE-free realization of the predictive observer is proposed. Simulation results are presented to demonstrate performance and convergence properties of the predictive observer in case of a wheeled mobile robot.

\end{abstract}

\section{INTRODUCTION}

The estimation problem in this paper is to determine attitude and position of a rigid body moving in a planar space. Attitude and position estimation are used in detecting and identifying faults \cite{c1} and effective attitude and position control of rigid bodies \cite{c2,c3}. The employed sensors in the landmark-based attitude and position estimation are divide into velocity sensors, such as rate Gyro and Doppler for angular and translational velocity readings, and charged-couple device (CCD) cameras for tracking terrain characteristics. Landmarks are points with known locations which can be observed by rigid body. Landmark-based attitude and position estimation have received considerable attention for the past decades \cite{c4,c5}. Except some sensors, such as laser detection and ranging (LADARs) or laser scanners which use light to determine the distance, landmark measurement is not usually available at the same time as it is perceived. This is either caused by long range observation, or complex computation, or both. Take CCD camera as an example of landmark measurement sensor, which is coupled with a video processing system. Due to computational burden of the image processing unit \cite{c6}, the time that takes to calculate the measurement vector from visual stream of information is not negligible and creates significant time-delay.

The effect of time-delay in sensor measurement, from the best of our knowledge, has been rarely dealt within the realm of rigid body attitude and position estimation \cite{c6}\nocite{c7,c8}-\cite{c9}. The existing works, are basically based on the Kalman filtering and its extensions. Smith predictors can also deal with delay in linear time-varying systems. However, both approaches do not guarantee asymptotic convergence of estimation error to zero due to intrinsic nonlinear and time-varying nature of attitude and position estimation problem. On the other hand, nonlinear observers \cite{c4,c10}\nocite{c12}-\cite{c14} 
 stand out as an important approach among a wide variety of estimation techniques. However, topological limitations on non-Euclidean spaces hamper achieving global stabilization. These limitations call for the relaxation from global to almost global stability, meaning that the region of attraction of the origin comprises all the state space except a nowhere dense set of measure zero \cite{c4,c15}.

In \cite{c6} authors presented an estimation method for combining measurements provided by inertial sensors (gyroscopes and accelerometers), global positioning system (GPS), and video system for unmanned aerial vehicle (UAV). The effect of data delay in video system is taken into account for attitude and position estimation. In \cite{c7,c8} the source of data delay was considered to be in the GPS sensor, where \cite{c7} proposed complementary Kalman filter and \cite{c8} proposed extended Kalman filter (EKF) to deal with estimation problem of interest.  A delay and dropout tolerant Kalman filter-based position, velocity, and acceleration estimation for aerial vehicles was proposed in \cite{c9} by fusing inertial and vision measurements. This work assumes vision measurement packets undergo delay and dropout due to image processing and wireless communication.

The contribution of this paper is the development of an attitude and position estimation algorithm in presence of time-delay in landmark measurements, while ensuring global and exponential convergence of estimation error. The estimation method is based on predictive observer of \cite{c16} which is further developed and employed for the case of attitude and position estimation problem. In the proposed technique, the overall attitude and position of the rigid body is described by an ordinary differential equation (ODE) in form of state affine systems. The delayed sensor measurement is formulated by a first-order Partial Differential Equation (PDE). The predictive observer, designed for the cascade of ODE-PDE systems, aims to predict and compensate delayed sensor measurement. Necessary and sufficient conditions for landmark configuration are presented to achieve asymptotic estimation of attitude and position. Convergence analysis are built on Lyapunov-Krasovskii functional. An upper-bound for sensor delay is derived which preserves convergence properties. Owing to simplified analysis, the observer is first designed under the ODE-PDE framework and then transformed into an implementable PDE-free realization.

The remainder of this paper is organized as follows. In Section II, we formulate our estimation problem of interest. In Section III, and IV the main result of the papers is presented. In Section V, we provide a PDE-free realization for implementation. 
A numerical example on wheeled mobile robot is presented in Section VI. Section VII concludes the paper.
\section{Problem statement and preliminaries}

\begin{figure}
\centering
\includegraphics[scale=1]{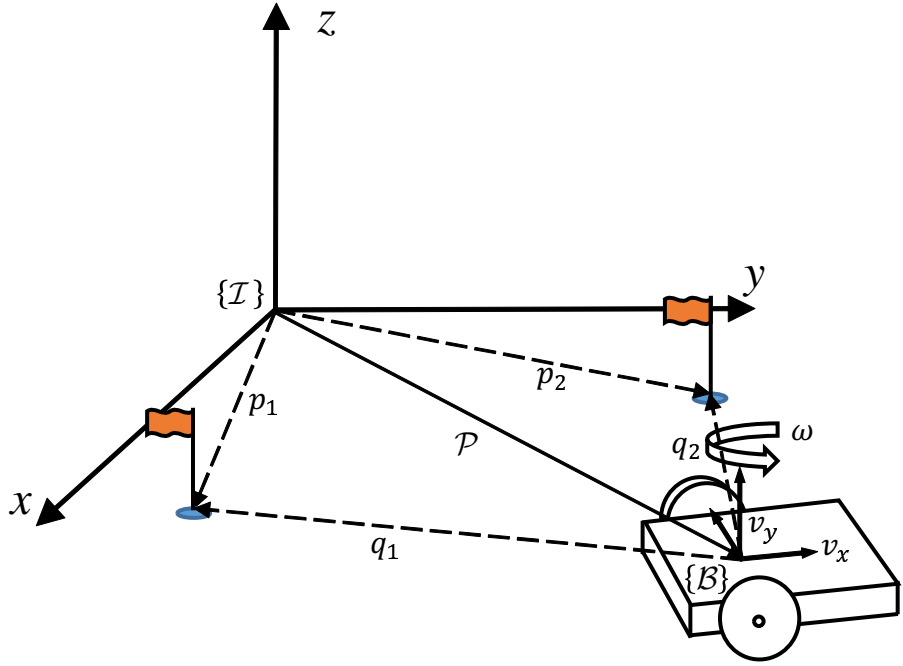}
\caption{Landmark-based attitude and position estimation.}
\label{F1}
\end{figure}

We denote $\left\{ {\cal I} \right\}$ as the inertial reference frame and $\left\{ {\cal B} \right\}$ as the body frame.  ${\cal P} \in {\mathbb R}^n$ denotes the position of the rigid body with respect to  $\left\{ {\cal I} \right\}$ and expressed in $\left\{ {\cal B} \right\}$. Attitude of the rigid body is represented by the rotation matrix ${\cal R} \in SO(n) = \left\{ {{\cal R} \in {{\mathbb R}^{n \times n}}\left| {\;\det ({\cal R}) = 1} \right.\;,\;{{\cal R}^{\top} }{\cal R} = I} \right\}$ where ${\cal R}^\top$ represents orientation of the body frame with respect to the inertial frame.
Rigid body attitude ${\cal R}$ and position ${\cal P}$ can be interpreted as an element of $ SE(n): = SO(n) \times {\mathbb R}^n$, which is represented by the matrix $$\left[ {\begin{array}{*{20}{c}} {\cal R}&{\cal P}\\ 0&1 \end{array}} \right] \in SE(n)$$ Thus, rigid body kinematics is described by
\begin{equation} \label{e1}
\left[ {\begin{array}{*{20}{c}}
{\dot {\cal R}}&{\dot {\cal P}}\\
0&1
\end{array}} \right] = \left[ {\begin{array}{*{20}{c}}
{ - S(\omega )}&v\\
0&0
\end{array}} \right]\left[ {\begin{array}{*{20}{c}}
{\cal R}&{\cal P}\\
0&1
\end{array}} \right]
\end{equation}
where $\omega  \in {\mathbb R}^n$ and $v \in {\mathbb R}^n$ denote angular and translational velocities of $\left\{ {\cal I} \right\}$ with respect to $\left\{ {\cal B} \right\}$, and expressed in $\left\{ {\cal B} \right\}$. We define the operator $S(\cdot)$ as a function from ${\mathbb R}^n$  to the space of skew-symmetric matrices $ S(n) = \left\{ {S \in {{\mathbb R}^{n \times n}}\left| \; \right. {S^{\top} } =  - S} \right\}$. Without loss of generality in planar motions,  namely $n=2$, angular velocity is a scalar variable. 
Landmark measurements indicated as ${q_i} \in {\mathbb R}^n,~i = 1, \cdots ,N $, are obtained through the sensors mounted on the rigid body which are capable of detecting and tracking terrain characteristics (such as CCD cameras). Also, we have
\[{q_i} = {\cal R}{p_i} - {\cal P}\]
where ${p_i}$ represents the location of $ {i^{th}} $ landmark with respect to the inertial frame $\left\{ {\cal I} \right\}$ and $ N $ is the number of landmarks. Fig. \ref{F1} depicts an example of rigid body planar motion in compliance with \eqref{e1}.

\subsection{Embedding $ SE(n)$ into the Euclidean Space}

Consider column stacking of ${\mathcal R}$ and ${\mathcal P}$ given by $ X= \begin{bmatrix} {\mathcal P}^\top & r_1^\top& \cdots&  r_n^\top \end{bmatrix}^\top \in {\mathbb R}^{{n^2} + n}$, where ${\cal R} = \begin{bmatrix} r_1& \cdots&  r_n \end{bmatrix},~ {r_j} \in {\mathbb R}^n,\;\;j = 1,...,n$. It is straightforward to see that equation (\ref{e1}) transforms into
\begin{equation} \label{e2}
\dot X(t) = A(\omega )X(t) + B(v)
\end{equation}
where $A(\omega ) =  - {\mathop{\rm diag}\nolimits} (S(\omega ), \ldots ,S(\omega )) \in {{\mathbb R}^{\left( {{n^2} + n} \right) \times \left( {{n^2} + n} \right)}}$ is a block diagonal matrix, and $ B(v) = {\left[ {\begin{array}{*{20}{c}} {{I_n}}&{{0_{n \times {n^2}}}} \end{array}} \right]^\top}v \in {{\mathbb R}^{{n^2} + n}}$ .
Similarly, by column stacking landmark measurements, the output equation $ {q_i} = {\cal R}{p_i} - {\cal P}$ can be formulated as $Y(t) = CX(t)$, where $Y = {\left[ {\begin{array}{*{20}{c}} {{y_1}^\top}& \cdots &{{y_n}^\top} \end{array}} \right]^\top} \in {{\mathbb R}^{nN}}$ and $$ C = \left[ {\begin{array}{*{20}{c}} { - {I_n}}&{{p_{11}}{I_n}}& \cdots &{{p_{1n}}{I_n}}\\  \vdots & \vdots & \ddots & \vdots \\
{ - {I_n}}&{{p_{N1}}{I_n}}& \cdots &{{p_{Nn}}{I_n}} \end{array}} \right] \in {{\mathbb R}^{nN \times ({n^2} + n)}}$$
with ${p_i} = {\left[ {\begin{array}{*{20}{c}} {{p_{i1}}}& \cdots &{{p_{in}}} \end{array}} \right]^{\top} } \in {{\mathbb R}^n}$.

Rigid body position and attitude kinematics can then be considered as a class of state affine systems described by
\begin{equation} \label{e3}
\begin{array}{l}
\dot X(t) = A(u)X(t) + B(u)\\
Y(t) = CX(t)
\end{array}
\end{equation}
where angular and linear velocities are lumped in $u(t) \subset {\cal D}$, where ${\cal D}$ is the set to which bounded inputs belong.
Taking into account the delayed stream of information from landmark measurement sensors, the estimation problem of interest in sequel is that of designing an observer for state affine system (3).

\subsection{Preliminaries}

\begin{df} \label{def1}  \cite{c17} When $A(u)$  satisfies the commutative property
\begin{equation} \label{eq:comproperty}
A(u(t))\left( {\int_{{t_0}}^t {A(u(\tau ))d\tau } } \right) = \left( {\int_{{t_0}}^t {A(u(\tau ))d\tau } } \right)A(u(t))
\end{equation}
for all $t,\;{t_0}$, then the state transition matrix associated with the system  $\dot X(t) = A(u)X(t)$ is given by
\begin{equation}  \label{e4}
\Phi (t,{t_0}) = {e^{\int_{{t_0}}^t {A(u(\tau ))d\tau } }}.
\end{equation}  \carrew
\end{df}
Properties pertinent to state transition matrix (\ref{e4}) can be enumerated as
\begin{itemize}
\item[1)] $\Phi ({t_0},{t_0}) = I$,
\item[2)] $\Phi (t,{t_0}) = \Phi (t,s)\Phi (s,{t_0})$,
\item[3)] $\frac{d}{{dt}}\Phi (t,{t_0}) = A(u(t))\Phi (t,{t_0})$,
\item[4)] ${\Phi ^{ - 1}}(t,{t_0}) = {\Phi ^\top}(t,{t_0}) = \Phi ({t_0},t)$,
\item[5)] $\Phi (t,{t_0})A(u(t)) = A(u(t))\Phi (t,{t_0})$.
\end{itemize}
Moreover, the transition matrix associated with $\dot {\cal R} =  - S(\omega ){\cal R}$ is ${\cal R}(t){{\cal R}^\top}({t_0})$. Therefore,  \eqref{e4} is equivalent to  
\begin{equation} \label{e5}
\Phi (t,{t_0}) = {\mathop{\rm diag}\nolimits} ({\cal R}(t){{\cal R}^\top}({t_0}), \cdots ,{\cal R}(t){{\cal R}^\top}({t_0}))\in {{\mathbb R}^{\left( {{n^2} + n} \right) \times \left( {{n^2} + n} \right)}}.
\end{equation}
Since ${\cal R}(\tau ){{\cal R}^\top}(t_0) \in SO(n)$, it follows that $\left\| {\Phi (\tau ,t)} \right\| = 1$.

\begin{rem} \label{remlimit}
The commutativity property in (\ref{eq:comproperty}) imposes limitation on the diversity of kinematics that can be fit into framework (\ref{e2}) and at the same time enjoy state transition matrix. Skew-symmetric matrices of dimension $2$ comply with this property. This makes systems with kinematics evolving in planar space, namely $SE(2)$, to be of practical interest to this theory. Meanwhile, to the best of authors knowledge, so far, no explicit state transition matrix is proposed for spatial kinematic evolution, namely $SE(3)$.
\end{rem} \carrew


\begin{df} \label{def2} The 2-norm of a vector is denoted by, $\|\cdot \| $. The $L_2$ norm of  matrix or vector functions (of the variable $x$) are denoted by ${\left\| {\;.\;} \right\|_{{L_2}\left[ {0,\;D} \right]}}$. In the sequel, the PDE state variable ${\cal U}(x,t)$ is a vector function of two variables $x$ and $t$, where $t$ is time and $x$ is a spatial variable that takes values in the interval $\left[ {0,\;D} \right]$. Therefore, $${\left\| {{\cal U}(t)} \right\|_{{L_2}\left[ {0,\;D} \right]}} = {\left( {\int_0^D {{{\cal U}^\top}(x,t){\cal U}(x,t)dx} } \right)^{1/2}}$$ which makes it a function of time variable $t$ \cite{c16}.  \carrew
\end{df}


\begin{lem} \label{lem1} \cite{c18,c19} \label{def3}  Consider the matrix differential equation
\begin{equation}\label{eq:Lyap}
\dot P(t) =  - \varepsilon P(t) - {A^{\top} }(u)P(t) - P(t)A(u) + {C^{\top} }{\rm{\Sigma }}C\
\end{equation}
where $\varepsilon \in {\mathbb R}_{> 0}$ is a positive constant and initial condition $P(0) \in {\mathbb R}^{n \times n}$ and parameter $ \Sigma \in {\mathbb R}^{n \times n}$ are symmetric positive definite (SPD) matrices. Then, there exist ${\beta _1},{\beta _2} \in {\mathbb R}_{> 0}$, such that $P(t)$ becomes a SPD matrix and satisfies the inequality $\,0 < {\beta _1}{I_n} \le P(t) \le {\beta _2}{I_n}$ for all time. Moreover, the lower bound of matrix differential equation (\ref{eq:Lyap}) is given by $ {\beta _1} = 2T{e^{ - \varepsilon T}}\underline{\lambda}({C^\top }{\rm{\Sigma }}C), \; \forall ~ t > {t_0} + T$ \cite{c18}, where $ \underline\lambda$ is the minimum eigenvalue of corresponding matrix. Parameter $T$ is a positive constant, so that, for all $t > {0}$
\[\int_t^{t + T} {{\Phi ^{\top} }(\tau ,t){C^{\top} }{\rm{\Sigma }}C\Phi (\tau ,t)d\tau }  \ge {\alpha}{I_n} > 0\]
where ${I_n}$ is the $ n \times n $ identity matrix and ${\alpha} = T\underline{\lambda}({\rm{\Sigma )}}\underline{\lambda}({C^{\top} }C{\rm{)}}$. \carrew
\end{lem}
Assuming that the rank of the matrix $ {C^{\top} }C $ is complete, the pair $(A(u),C)$ is uniformly completely observable and hence $\beta_1>0$.


\begin{lem} \label{young} For any  ${\cal X},\;{\cal Y} \in {\mathbb R}^{n}$ and any scalar $\kappa \in {\mathbb R}_{>0}$ , one has
\begin{equation} \label{e6}
2{{\cal X}^{\top} }{\cal Y} \le \kappa {{\cal X}^{\top} }{\cal X} + {\kappa ^{ - 1}}{{\cal Y}^{\top} }{\cal Y}.
\end{equation} \carrew
\end{lem}

\section{Predictor observer }
Consider the following state affine system
\begin{equation} \label{e7}
\begin{array}{l}
\dot X(t) = A(u)X(t) + B(u)\\
Y(t) = CX(t - D)
\end{array}
\end{equation}
where $X(t)$, $u(t)$ and $Y(t)$ are the state, input, and output of the system, respectively. Also, $ A(u),\;B(u),\;C $ are uniformly bounded matrices of compatible dimensions. We assume that the output is delayed by $D$ unit of time, which is constant and known. Let the predictive observer for system (\ref{e7}) be given by
\begin{equation}\label{eq:Observer}
\begin{array}{ll}
\dot{\hat {X}}(t) = A(u)\hat X(t) + B(u) + {e^{\int_0^D {A(u(\tau ))d\tau } }}{P^{ - 1}}(t){C^{\top} }{\rm{\Sigma }}\left( {Y(t) - \hat Y(t)} \right) \\
\hat Y(t) = C\hat X(t - D) + C\int_{t - D}^t {{e^{\int_0^{t - \theta } {A(u(\tau ))d\tau } }}{P^{ - 1}}(\theta ){C^{\top} }{\rm{\Sigma }}\left( {Y(\theta ) - \hat Y(\theta )} \right)d\theta } \\
\dot P(t) =  - \varepsilon P(t) - {A^{\top} }(u)P(t) - P(t)A(u) + {C^{\top} }{\rm{\Sigma }}C 		
\end{array}
\end{equation}
where $\hat X(t)$ and $\hat Y(t)$ are observer state and output, respectively. The rest of the notations are elaborated on \textit{Lemma}~\ref{lem1} and \textit{Definition}~\ref{def1}.


\begin{rem} \label{remphi} Observer output in (\ref{eq:Observer}) have a distributed delay integral feedback term. This term is in form of a general convolution integral function given by
\[\int_{t - D}^t {\Phi (t - \theta ,0)f(\theta )d\theta }, \]
where $ \Phi (t - \theta ,0) = {e^{\int_0^{t - \theta } {A(u(\tau ))d\tau } }} $ and $ f(t) = {P^{ - 1}}(t){C^{\top} }{\rm{\Sigma }}\left( {Y(t) - \hat Y(t)} \right) $.
The distributed delay integral term depending on the the output estimation error ${Y(t) - \hat Y(t)}$, is prominent feature of the predictor based observer and ensures exponential convergence of estimation error to zero. 
However, in most standard state observers for time-delay systems, only a pure output error term is presented and the convergence is normally asymptotic. \carrew
\end{rem}


\section{Observer Synthesis}

To simplify convergence analysis, we model the output equation in (\ref{e7}) by PDE
\begin{equation}  \label{e9}
\begin{array}{l}
{{\cal U}_t}(x,t) = {{\cal U}_x}(x,t)\\
{\cal U}(D,t) = CX(t)\\
Y(t) = {\cal U}(0,t)
\end{array}
\end{equation}
where the delayed state ${\cal U}(x,t)$ depends on the time variable $t$ and the spatial variable $x$. Variable $x$ assumes values in the interval $\left[ {0,\;D} \right]$.
It can be verified that the solution to this transport PDE equation is
\begin{equation} \label{defu}
{\cal U}(x,t) = CX(t + x - D)
\end{equation}
Therefore, at the boundary condition $x = 0$, we have the delayed state ${\cal U}(0,t) = CX(t - D)$, which is equivalent to system output.
The entire delayed system can then be represented as interconnection of ODE and PDE
\begin{equation} \label{e10}
\begin{array}{l}
\dot X(t) = A(u)X(t) + B(u)\\
{{\cal U}_t}(x,t) = {{\cal U}_x}(x,t)\\
Y(t) = {\cal U}(0,t)
\end{array}
\end{equation}
Now, we propose the predictive observer in the ODE-PDE form as
\begin{equation}    \label{e11}
 \begin{array}{l}
\dot{\hat {X}}(t) = A(u)\hat X(t) + B(u) + \Phi (D,0){P^{ - 1}}(t){C^{\top} }{\rm{\Sigma }}\left( {Y(t) - \hat {\cal U}(0,t)} \right)\\
{{\hat {\cal U}}_t}(x,t) = {{\hat {\cal U}}_x}(x,t) + C\Phi (x,0){P^{ - 1}}(t){C^{\top} }{\rm{\Sigma }}\left( {Y(t) - \hat {\cal U}(0,t)} \right)\\
\hat {\cal U}(D,t) = C\hat X(t)
\end{array}
\end{equation}
where $\Phi(t,0)$ is the transition matrix (\ref{e4}) and $P(t)$ is given by the matrix differential equation in \textit{Lemma}~\ref{lem1}. A block diagram of the proposed observer is provided in Fig. 2. Note that the above ODE-PDE representation of the observer is for analysis purpose. Real-time implementation of the observer is always based on (\ref{eq:Observer}).
\begin{figure}[thpb] \label{F2}
\centering
\includegraphics[scale=0.8]{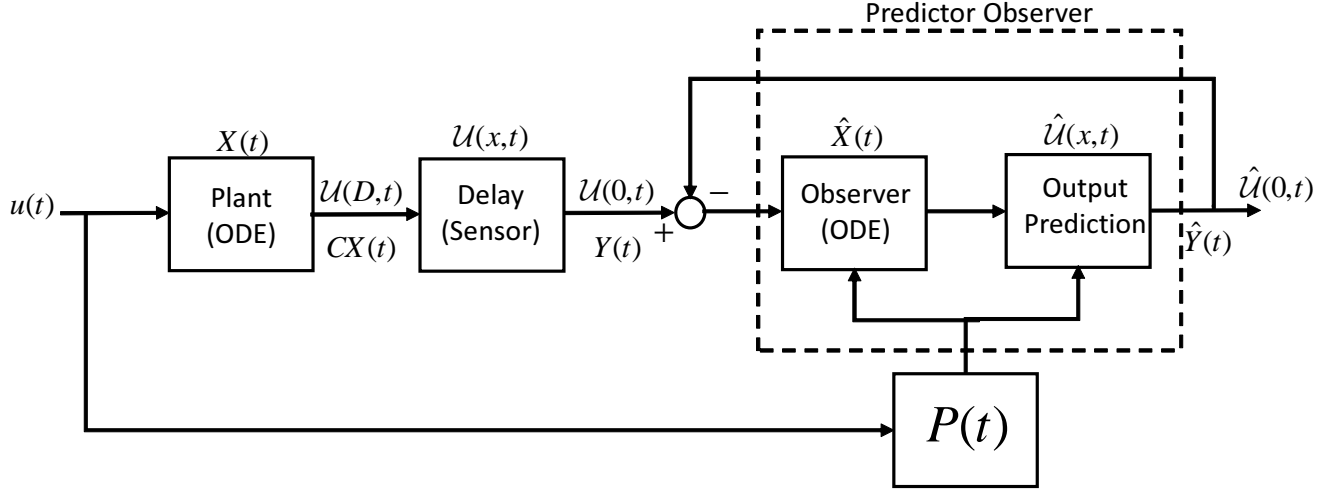}
\caption{Block diagram of the predictive observer in presence of sensor delay $D$.}
\end{figure}

Next, we define the estimation error variables by
$$\tilde X(t) = X(t) - \hat X(t)$$
$$\tilde {\cal U}(x,t) = {\cal U}(x,t) - \hat {\cal U}(x,t)$$
By virtue of (\ref{defu}), the term $\tilde {\cal U}(x,t)$ can be regarded as delayed sensor estimation error.
The following error dynamics are resulted from (\ref{e10}) and (\ref{e11}).
\begin{equation}  \label{e12}
 \begin{array}{l}
\dot{\tilde {X}}(t) = A(u)\tilde X(t) - \Phi (D,0){P^{ - 1}}(t){C^{\top} }{\rm{\Sigma }}\left( {Y(t) - \hat {\cal U}(0,t)} \right)\\ {{\tilde {\cal U}}_t}(x,t) = {{\tilde {\cal U}}_x}(x,t) - C\Phi (x,0){P^{ - 1}}(t){C^{\top} }{\rm{\Sigma }}\left( {Y(t) - \hat {\cal U}(0,t)} \right)\\
\tilde {\cal U}(D,t) = C\tilde X(t)
\end{array}
\end{equation}
Global and exponential convergence of the error dynamics (\ref{e12}) is investigated in sequel.
We define the composite estimation error by \cite{c16}
\begin{equation}   \label{error}
\tilde {\cal W}(x,t) = \tilde {\cal U}(x,t) - C\Phi (x,D)\tilde X(t)
\end{equation}
where $\tilde {\cal W}(x,t)$ includes the estimation error $\tilde X(t)$ and the delayed output estimation error $\tilde {\cal U}(x,t)$. Introduction of the composite error is to further simplify and clarify the convergence analysis.
Differentiating $\tilde {\cal W}(x,t)$ with respect to $x$ and $t$ with appropriate replacement from (\ref{e12}), yields
\begin{equation}   \label{e14}
{\tilde {\cal W}_{\;t}}(x,t) - {\tilde {\cal W}_{\;x}}(x,t) = C\left( {A(u(x)) - A(u(t))} \right)\Phi (x,D)\tilde X(t)
\end{equation}
In light of the above transformation and the boundary condition $\tilde {\cal W}(D,t) = \tilde {\cal U}(D,t) - C\tilde X(t) = 0$, the observer error dynamics (\ref{e12}) becomes
\begin{equation}    \label{e15}
 \begin{array}{l}
\dot{\tilde {X}}(t) = \left( {A(u) - \Phi (D,0){P^{ - 1}}(t){C^{\top} }{\rm{\Sigma }}C\Phi (0,D)} \right)\tilde X(t) - \Phi (D,0){P^{ - 1}}(t){C^{\top } }{\rm{\Sigma }}\tilde {\cal W}(0,t)\\
{{\tilde {\cal W}}_t}(x,t) = {{\tilde {\cal W}}_x}(x,t) + C\left( {A(u(x)) - A(u(t))} \right)\Phi (x,D)\tilde X(t)\\
\tilde {\cal W}(D,t) = 0
\end{array}
\end{equation}
The following theorem is the main result of this section.

\begin{theo} \label{th1} The attitude and position observer (\ref{e11}) guarantees that $\lim_{t \to +\infty} \tilde X(t) =0\ $ and $\lim_{t \to +\infty} \tilde{\cal U}(x,t)=0, \forall x \in [0,D] $. More specifically, the observer error equation (\ref{e15}) is exponentially stable in the sense of the norm
\begin{equation} \label{norm}
{\left( {{{\| {\tilde X(t)} \|}^2} + \int_0^D {{{\tilde {\cal U}}^\top}(x,t)\tilde {\cal U}(x,t)dx} } \right)^{1/2}}
\end{equation}
provided that the following two assumptions hold:
\begin{itemize}
\item[1)] The known and constant time-delay satisfies $0 < D < D_{max}$.
\item[2)] The observer gain is chosen to satisfy $0 <\varepsilon_{min} < \varepsilon < \varepsilon_{max}$.\\
where $\varepsilon_{min},\varepsilon_{max},D_{max}$ are given in the proof.\\
{\it Proof:} In order to preserve continuity, a proof of this result is provided in Appendix. 
\end{itemize} \carrew
\end{theo}

Both the maximum tolerable delay and the observer gain directly depend on design parameters and landmark configuration, namely matrix $C$. For a given sensor delay, the observer gain assumes its admissible values in an open interval determined in Assumption 2 of the \textit{Theorem}~\ref{th1}. Note that the landmarks are not all collinear (on a straight line), otherwise matrix $C$ would be singular and consequently no specific interval for the observer gain $\varepsilon $ could be found to establish the observation convergence.

The following Theorem provides characteristics of exponential convergence of the norm (\ref{norm}).
\begin{theo} \label{th2}  An exponentially decaying upper bound for the estimation error norm is given by
\begin{equation}   \label{e18}
{\left({\| {\tilde X(t)} \|^2} + \int_0^D {{{\tilde {\cal U}}^\top}(x,t)\tilde {\cal U}(x,t)dx}\right)^{1/2}}  \le \sqrt {\frac{{{\varphi _2}{\psi _2}}}{{{\varphi _1}{\psi _1}}}} {e^{ - \frac {\mu}{2} t}}{\left( {{{\| {\tilde X(0)} \|}^2} + \int_0^D {{{\tilde {\cal U}}^\top}(x,0)\tilde {\cal U}(x,0)dx} } \right)^{1/2}}
\end{equation} \carrew
\end{theo}
{\it Proof:} In order to preserve continuity, a proof of this result is provided in Appendix.

A PDE-free realization of the observer (\ref{e11}) is provided in the next section.

\section{Observer Implementation}

In this section an equivalent representation of predictive observer (11) out of ODE-PDE form is derived. This representation is of importance in implementation and further understanding of the predictive observer.
Taking the Laplace transform ${\cal L}$, from the PDE equation in (\ref{e11}), and knowing that $\hat {\cal U}(x,0) = 0$, yields
\[\begin{array}{l}
s\hat {\cal U}(x,s) = \frac{d}{{dx}}\hat {\cal U}(x,s) + C\Phi (x,0){\cal L}\left\{ {{P^{ - 1}}(t){C^{\top} }{\rm{\Sigma }}\left( {Y(t) - \hat Y(t)} \right)} \right\}\\
\hat {\cal U}(0,s) = \hat Y(s)
\end{array}\]
The solution to this first-order ODE in terms of $x$ is
\[\hat {\cal U}\,(x,s) = \hat Y(s){e^{sx}} - \int_0^x {{e^{s(x - \eta )}}C} \Phi (\eta ,0){\cal L}\left\{ {{P^{ - 1}}(t){C^{\top} }{\rm{\Sigma }}\left( {Y(t) - \hat Y(t)} \right)} \right\}d\eta \]
Inserting the boundary condition $ \hat {\cal U}(D,s) = CX(s) $ in the above equation, we have 
\[\hat Y(s) = C\hat X(s){e^{ - sD}} + \int_0^D {{e^{ - s\eta }}C} \Phi (\eta ,0){\cal L}\left\{ {{P^{ - 1}}(t){C^{\top} }{\rm{\Sigma }}\left( {Y(t) - \hat Y(t)} \right)} \right\}d\eta \]
Finally, after taking the inverse Laplace transform and a change of variable $\theta  = t - \eta $, we obtain
\[\hat Y(t) = C\hat X(t - D) + C\int_{t - D}^t {\Phi (t - \theta ,0){P^{ - 1}}(\theta ){C^{\top} }{\rm{\Sigma }}\left( {Y(\theta ) - \hat Y(\theta )} \right)d\theta } \]
Thus, the observer representation in terms of the output is given by
\begin{equation} \label{e20}
\begin{aligned}
&\dot{\hat {X}}(t) = A(u)\hat X(t) + B(u) + \Phi (D,0){P^{ - 1}}(t){C^{\top} }{\rm{\Sigma }}\left( {Y(t) - \hat Y(t)} \right)\\
&\hat Y(t) = C\hat X(t - D) + C\int_{t - D}^t {\Phi (t - \theta ,0){P^{ - 1}}(\theta ){C^{\top} }{\rm{\Sigma }}\left( {Y(\theta ) - \hat Y(\theta )} \right)d\theta }
\end{aligned}
\end{equation}
The predictive observer (\ref{e20}) is a PDE-free realization of observer (\ref{e11}) that involves a distributed delay integral feedback term in observer output.

\section{Numerical example: Wheeled Mobile robot}

We consider a wheeled mobile robot, depicted in Fig. 1, which complies with the class of state affine systems for which the predictive observer is designed. 
Owing to planar motion of the mobile robot, its rotation is around a single axes, namely the $z$ axis. Hence, the angular velocity $ \omega (t) $ is a bounded scalar value. Furthermore, the employed wheeled robot can not have any displacement along the axis perpendicular to their wheels. This makes its translational velocity vector (expressed in body frame) as  $ v = [{v_x} ~ 0]^{\top} $.
\begin{figure}[thpb]  \label{F3}
\centering
\includegraphics[scale=0.7]{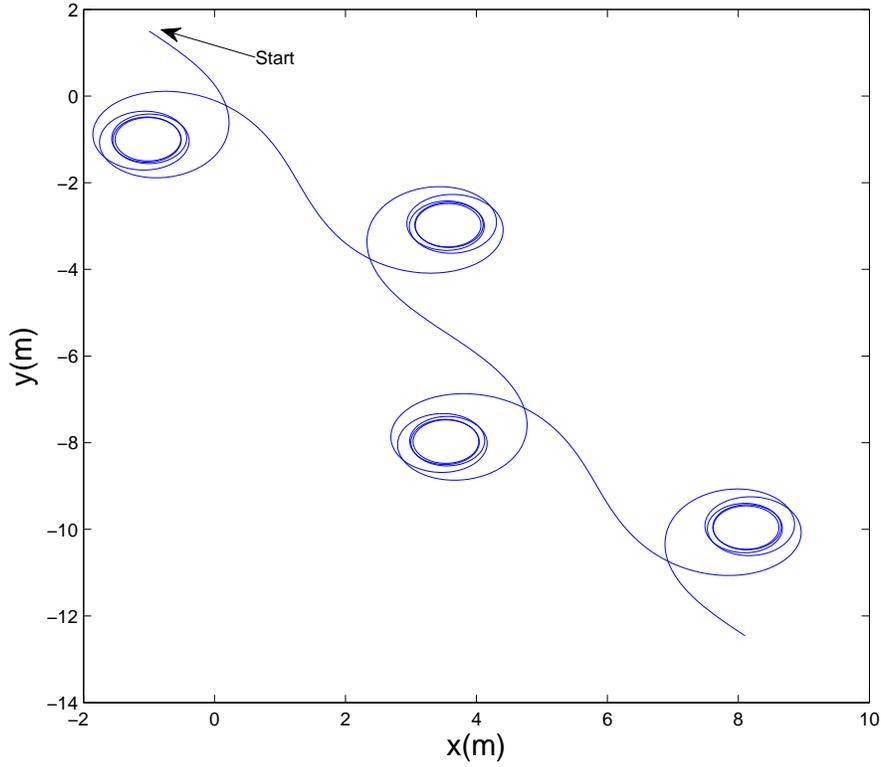}
\caption{Wheeled mobile robot planar path. }
\end{figure}

We place our landmarks physically at locations $ {p_1} = [1~3]^{\top}m$ and ${p_2} = [3~ 1]^{\top}m$. Adding these landmark locations we find the third location to be $ {p_3} = [4 ~ 4]^{\top}m$. After embedding the robot kinematics from $SE(2)$ into ${\mathbb R}^6$, the attitude and position dynamics take the form of state affine systems (\ref{e7}). Therefore, the pertinent matrices $A(u)$, $B(u)$, and $C$ are given by
$$A(u) = {\mathop{\rm diag}\nolimits} (S(\omega ),S(\omega ),S(\omega )), \quad S(\omega ) = \left[ {\begin{array}{*{20}{c}} 0&{ - \omega }\\ \omega &0 \end{array}} \right], \quad B(u) = {\left[ {\begin{array}{*{20}{c}}{{v_x}}&0&0&0&0&0 \end{array}} \right]^{\top} } $$ $$ \; C = \left[ {\begin{array}{*{20}{c}}
{ - {I_2}}&{{I_2}}&{3{I_2}}\\ { - {I_2}}&{3{I_2}}&{{I_2}}\\ { - {I_2}}&{4{I_2}}&{4{I_2}} \end{array}} \right] $$
The planar motion of the robot in 100 seconds is depicted in the $X-Y$ plane and shown in Fig. 3. The angular and linear velocities are given by $\omega (t) = 2\sin (0.04\pi t)\;{\rm{rad}}/{\rm{s}}$ and ${v_x} = 1\;{\rm{m}}/{\rm{s}} $. Furthermore, The initial values are arbitrarily selected as $ X(0) = {\left[ {\begin{array}{*{20}{c}}{ - \frac{5}{{\sqrt 2 }}}&{\frac{1}{{\sqrt 2 }}}&{\frac{{\sqrt 2 }}{2}}&{\frac{{\sqrt 2 }}{2}}&{ - \frac{{\sqrt 2 }}{2}}&{\frac{{\sqrt 2 }}{2}}\end{array}} \right]^{\top}}$. In order to assess the performance of the proposed predictive observer (\ref{e20}), we consider a standard state observer of the form \cite{c20}
\begin{equation}   \label{e21}
\begin{array}{l}
\dot{\hat {X}}(t) = A(u)\hat X(t) + B(u) + {P^{ - 1}}(t){C^{\top} }{\rm{\Sigma }}\left( {Y(t) - \hat Y(t)} \right)\\
\hat Y(t) = C\hat X(t - D)\\\dot P(t) =  - \varepsilon P(t) - {A^{\top} }(u)P(t) - P(t)A(u) + {C^{\top} }{\rm{\Sigma }}C
\end{array}
\end{equation}
and compare it with the proposed observer. In the standard observer we set $\varepsilon $ and ${\rm{\Sigma }}$ similar to the predictive observer. Landmark measurements are assumed to be available after $D$ unit of time. Furthermore, both observers assume the arbitrary initial values $P(0) = 0.5{I_{6 \times 6}}$ and ${\rm{\Sigma  = 0}}{\rm{.5}}{I_{6 \times 6}} $.
\begin{figure}[thpb] \label{F4}
\centering
\includegraphics[scale=0.9]{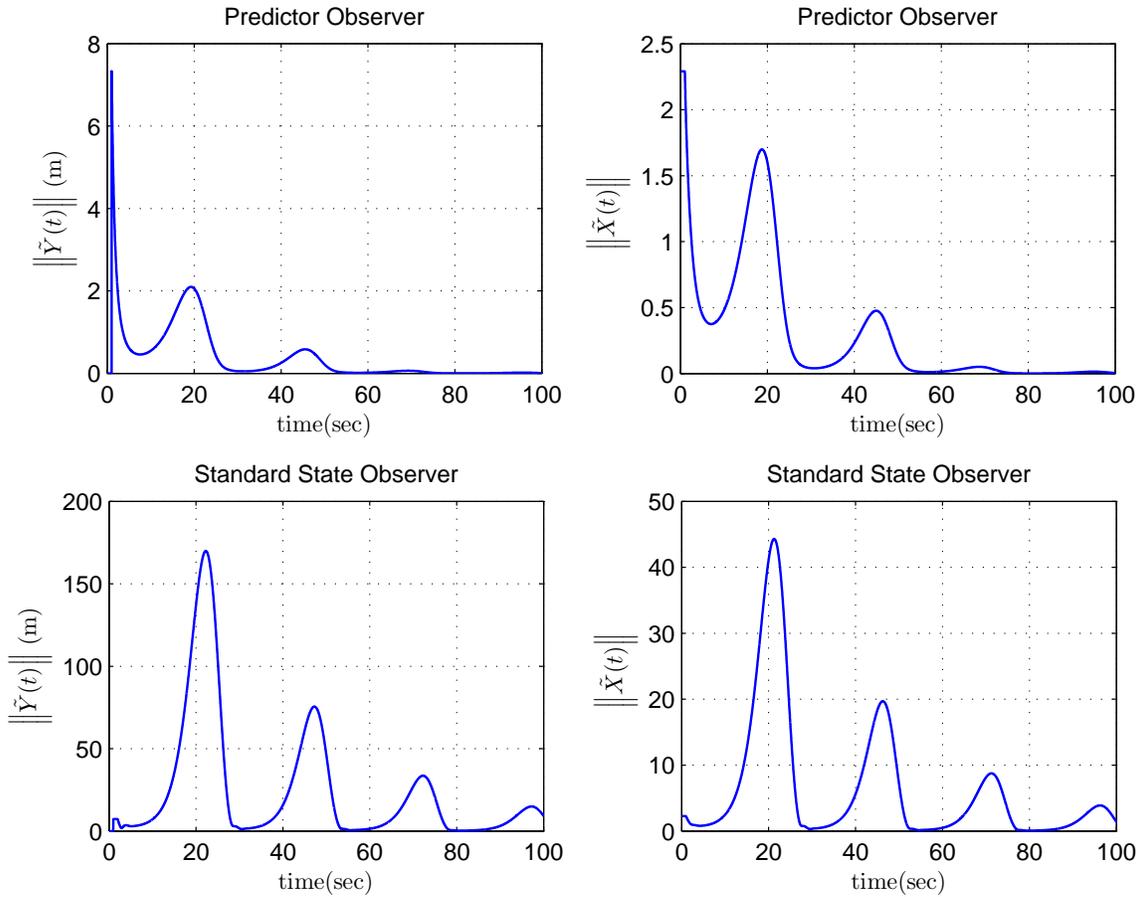}
\caption{Comparison of the predictive and the standard observer for $D=1$ and $ \varepsilon=0.6 $. }
\end{figure}

The observer gain $ \varepsilon $ is  selected according to assumption 2 in Theorem 1. The pertinent range for $ \varepsilon $ is found from simulations for different delays. From numerical simulations it is observed that the standard observer (\ref{e21}) does not converge for delays of larger than 1.04 second, whereas the predictive observer (\ref{e20}) is still convergent for delays of shorter than 1.5 second. In simulations for delays $D=1.1$, $D=1.2$, $D=1.3$, $D=1.4$, and $D=1.5$, the lower bound of $ \varepsilon $ is found to be $0.2$, $1.6$, $2.1$,$5$ and $23$, respectively.
In Fig. 4 both the predictive and the standard observer are compared for $D=1$ and $ \varepsilon=0.6 $. Since the error of the standard observer grows arbitrary large for delays of larger than 1 second, the results of this observer is eliminated hereafter. The predictive observer performs satisfactorily for constant delays $D=1.2$ and  $D=1.4$ and different values of $ \varepsilon$ as illustrated in Fig. 5. Furthermore, in Fig. 6 the effect of variation in delay is illustrated for a fixed observer gain.
\begin{figure}[thpb] \label{F5}
\centering
\includegraphics[scale=0.9]{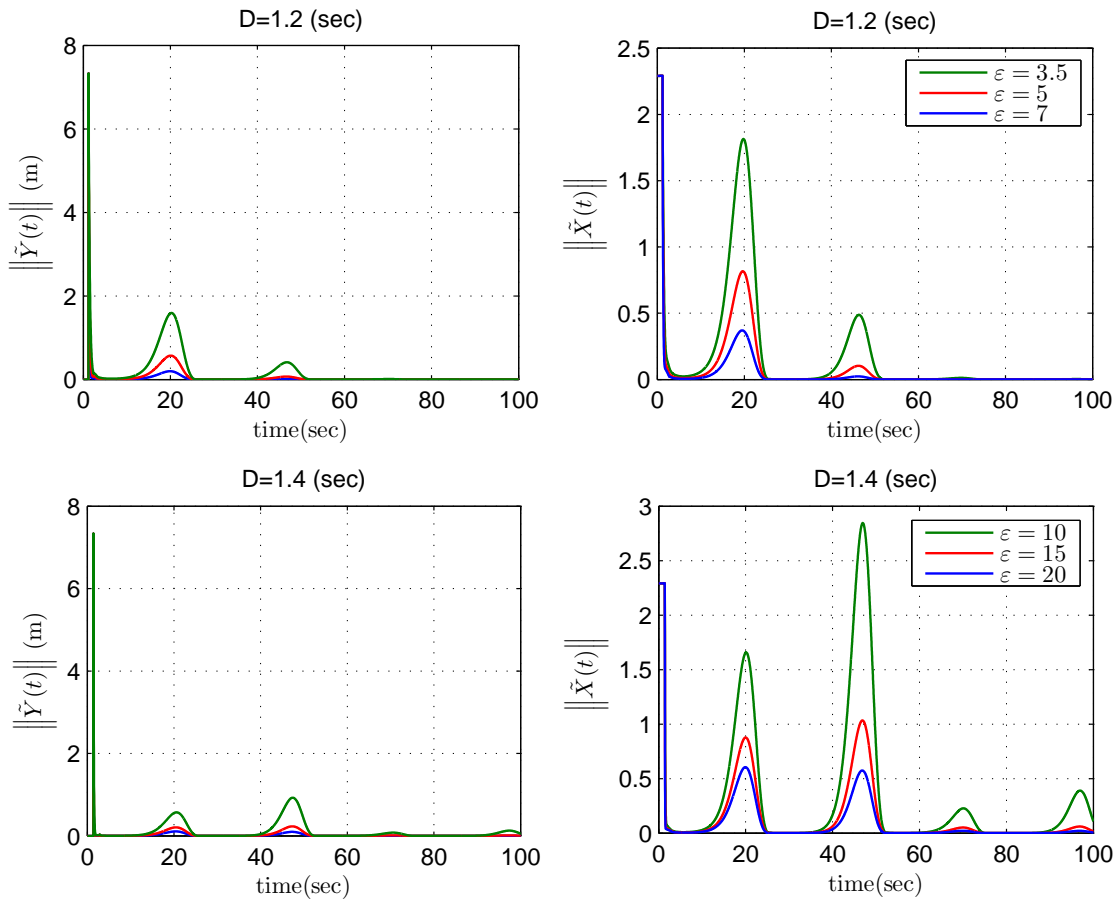}
\caption{Performance of the predictive observer for $D=1.2$, $D=1.4$, and different values of the observer gain.}
\end{figure}

\begin{figure}[thpb] \label{F6}
\centering
\includegraphics[scale=0.9]{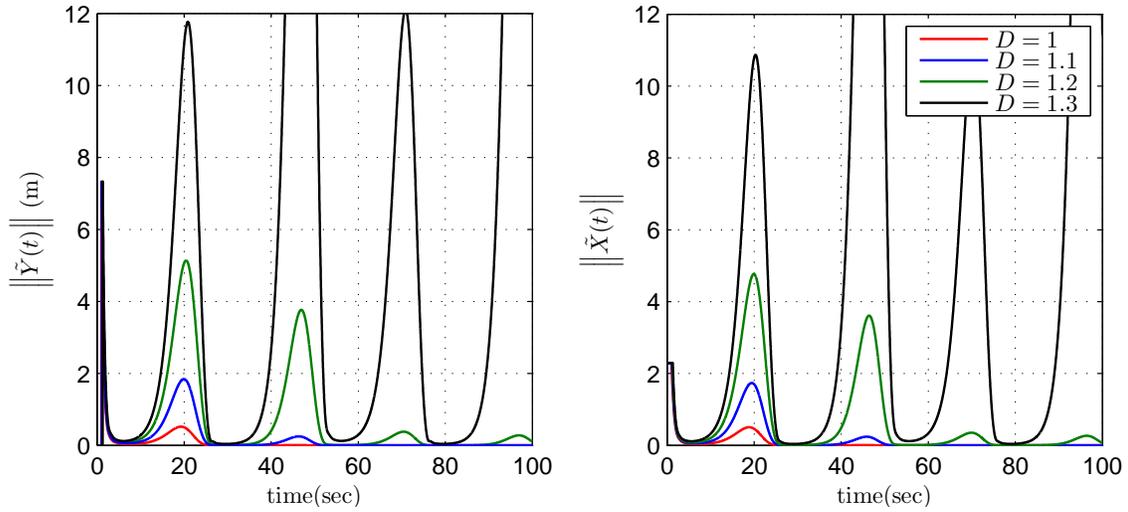}
\caption{Performance of the predictive observer for $ \varepsilon=2 $ and different values of delay.}
\end{figure}
From Fig. 4 and 5, it can be understood that the predictive observer (\ref{e20}) outperforms the standard state observer (\ref{e21}); in the sense that it handles larger sensor delays, enjoys faster convergence, and ensures smaller error.

\begin{figure}[thpb]
\centering
\includegraphics[scale=0.9]{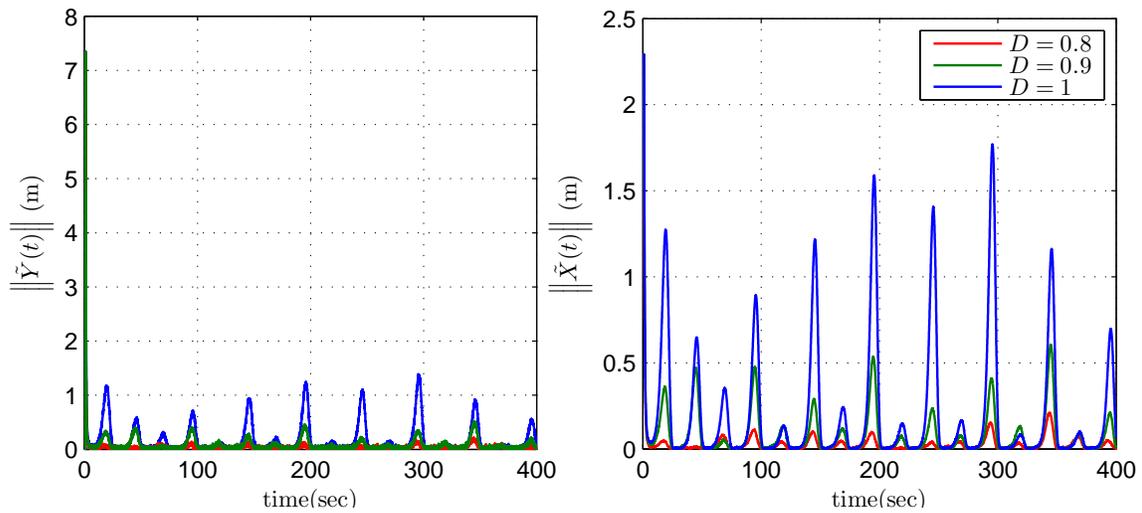}
\caption{Performance of the predictive observer in presence of noise and for $\varepsilon = 2$ and different values of delay.}
\label{F7}
\end{figure}

 \subsection{Sensor noise} Noise in sensors is considered on linear velocity readings and landmark measurements. In particular, additive, zero-mean, white Gaussian noise is taken into account, with standard deviation of $4 ~cm/s$ (4 percent) for linear velocity and $0.04~m$ for the landmark measurements.

As we can see in Fig~\ref{F7}, $\tilde Y(t)$ is noisier in comparison to $\tilde X(t)$, while error bound in $\tilde X(t)$ is bigger than that of $\tilde Y(t)$. In presence of noise the observer is sensitive with respect to output delay. This means the steady state error for both state and output has a finite bound, while this bound increases by the growth in sensor delay. We observe that even in presence of noise, the predictive observer demonstrates a plausible behavior, though the maximum tolerable delay in output sensor decreases.

\begin{figure}[thpb]
\centering
\includegraphics[scale=0.9]{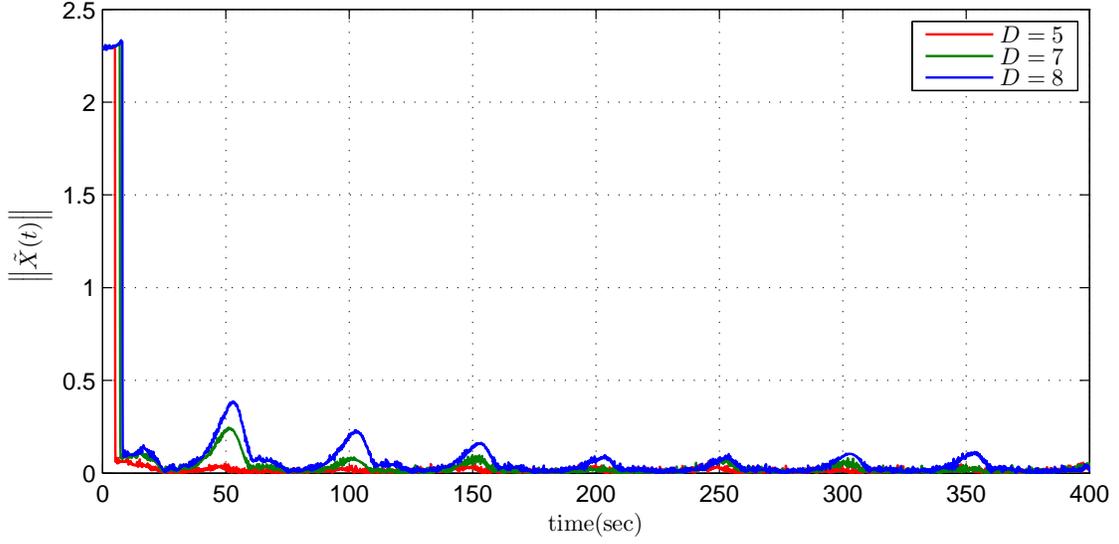}
\caption{Performance of the predictive observer in presence of noise and for $\varepsilon = 8$ and different values of delay.}
\label{F8}
\end{figure}

To incorporate noise, as a realistic phenomenon, a realistic angular velocity is required as well, for instance, lets say $\omega=0.4\sin(0.04\pi t)~rad/sec$. From definition of $\gamma$ in (\ref{eq:gamma}) and the maximum tolerable delay in (\ref{eq:Dmax}) we see that decrease in angular velocity (equivalently parameter $\gamma$) leads to a bigger $D_{max}$. This is corroborated by simulation with adopted angular velocity. The noise-free observer in this case can tolerate output delay up to $D_{max}=8.7$; Whereas, in presence of noise Fig~\ref{F8}, $\tilde X(t)$ tend to grow larger by increase in output time-delay $D$. This example better elucidates the sensitivity and performance degradation of predictive observer in presence of sensor noise for large sensor delays.

\section{conclusions }

This paper presents an attitude and position observer based on landmark measurements and velocity readings. The attitude and position estimations are obtained from a globally exponential stable predictive observer combining the measurements from velocity sensors together with landmark measurements. It is assumed landmark sensors have time-delay in measuring landmark's position. Upper bound of the time-delay for which the observer converges, is calculated. Simulation results confirm advantages of the predictive observer over a standard state observer. As a possible future line of research, it is interesting to investigate the same problem under state dependent delay.


\section{appendix}

\subsection{Proof of {\it Theorem 1}}

{\it Proof:}
Consider the Lyapunov-Krasovskii functionals
\begin{equation} \label{e16}
\begin{aligned}
&V(t) = {V_1}(t) + {V_2}(t) \\
&{V_1}(t) = {{\tilde X}^{\top} }(t){\Phi ^{\top} }(0,D)P(t)\Phi (0,D)\tilde X(t)  \\
&{V_2}(t) = \rho \int_0^D {(1 + x){{\tilde {\cal W}}^{\top} }(x,t)\tilde {\cal W}(x,t)dx} 	
\end{aligned}
\end{equation}
where $\rho $ is a positive scalar to be chosen. By virtue of (\ref{e16}), it can be inferred that
\begin{equation} \label{e17}
\begin{aligned}
& ~~~~~~~~~~~~~~~~~~~~~~~~{\beta _1}{\left\| {\tilde X(t)} \right\|^2} \le {V_1}(t) \le {\beta _2}{\left\| {\tilde X(t)} \right\|^2}  \\
&\rho \int_0^D {{{\tilde {\cal W}}^{\top} }(x,t)\tilde {\cal W}(x,t)dx}  \le {V_2}(t) \le \rho (1 + D)\int_0^D {{{\tilde {\cal W}}^{\top} }(x,t)\tilde {\cal W}(x,t)dx}
\end{aligned}
\end{equation}
where $\beta_1,~\beta_2$ are first appeared in \textit{Lemma}~\ref{lem1}, Taking time differentiation of the functionals in (\ref{e16}) and substituting from (\ref{e15}), yields
\begin{align*}
{{\dot V}_1}(t) = &  - {{\tilde X}^{\top} }(t){\Phi ^{\top} }(0,D)\big( {\varepsilon P(t) + {C^{\top} }{\rm{\Sigma }}C} \big)\Phi (0,D)\tilde X(t) - 2\tilde X{(t)^{\top} }{\Phi ^{\top} }(0,D){C^{\top} }{\rm{\Sigma }}\tilde {\cal W}(0,t)\\
{{\dot V}_2}(t) = &  - \rho {{\tilde {\cal W}}^{\top} }(0,t)\tilde {\cal W}(0,t) - \rho \int_0^D {{{\tilde {\cal W}}^{\top} }(x,t)\tilde {\cal W}(x,t)dx} \\
& + \rho \int_0^D {2(1 + x){{\tilde {\cal W}}^{\top} }(x,t)C\left( {A(u(x)) - A(u(t))} \right)\Phi (x,D)\tilde X(t)dx}
\end{align*}
 In light of properties 2 and 4 in  Definition \ref{def1}, and the lower bound of $ P(t) $ given in Lemma \ref{def3} and the inequality (\ref{e6}), we have
\begin{align*}
{{\dot V}_1}(t) \le & - \varepsilon {\beta _1}{\| {\tilde X(t)} \|^2} + \left( {{\kappa _1}\bar\lambda ({C^{\top} }{{\rm{\Sigma }}^2}{\rm{C}}) - \underline\lambda ({C^{\top} }{\rm{\Sigma }}C)} \right){\| {\tilde X(t)} \|^2} + \kappa _1^{ - 1}{\| {\tilde {\cal W}(0,t)} \|^2}\\
{{\dot V}_2}(t) \le & - \rho {\| {\tilde {\cal W}(0,t)} \|^2} - \rho \int_0^D {{{\tilde {\cal W}}^{\top} }(x,t)\tilde {\cal W}(x,t)dx} + \rho  \Big[ {(1 + D)\big( {{\kappa _2}\int_0^D {{{\tilde {\cal W}}^{\top} }(x,t)\tilde {\cal W}(x,t)dx} } \big)  } \\
& +  {\kappa _2^{ - 1}\bar \lambda ({C^{\top} }C) \| {\tilde X(t)} \|}^2 \int_0^D {(1 + x){{\| {\Phi (x,D)} \|}^2}} {{\| {A(u(x)) - A(u(t))} \|}^2}dx  \Big ]
\end{align*}
Since $ {\cal R}(x){{\cal R}^\top}(D) \in SO(n) $ it follows from (\ref{e5}) that $ \| {\Phi (x,D)} \| = 1 $. Moreover,
\begin{equation}
\gamma  := \mathop {\sup }\limits_{u \in {\cal D}} \left\{ {\left\| {A(u(x)) - A(u(t))} \right\|_{{L_2}\left[ {0,\;D} \right]}^2} \right\}
\label{eq:gamma}
\end{equation}
Choosing $ \rho  = \kappa _1^{ - 1} $  in the Lyapunov function, implies

\begin{equation} \label{eq18}
\dot V(t) \le  - {\delta _1}{\| {\tilde X(t)} \|^2} - \rho {\delta _2}\| {\tilde {\cal W}(t)} \|_{{L_2}\left[ {0,\;D} \right]}^2
\end{equation}
where
\[{\delta _1} = \varepsilon {\beta _1} + \underline\lambda {({C^{\top} }{\rm{\Sigma }}C)} - {\kappa _1}\bar\lambda ({C^{\top} }{{\rm{\Sigma }}^2}{\rm{C}}) - D(\frac{D}{2} + 1)\gamma \kappa _1^{ - 1}\kappa _2^{ - 1}\bar\lambda ({C^{\top} }C)\]
\[{\delta _2} = 1 - (1 + D){\kappa _2}\]
In order to make $ {\delta _2} > 0 $, the parameter $ {\kappa _2} $ must be chosen such that $ {\kappa _2} < \frac{1}{{1 + D}} $. An equivalent expression can be derived as $ 1 + D < \kappa _2^{ - 1} $. Eliminating $\kappa _2^{ - 1} $ from $ {\delta _1} $, yields
\begin{equation} \label{epsilon}
\varepsilon T{e^{ - \varepsilon T}} \ge \frac{1}{2}\left[ {\frac{{D(D + 1)(D + 2)\gamma \kappa _1^{ - 1}\bar\lambda ({C^{\top} }C) + 2{\kappa _1}\bar\lambda ({C^{\top} }{{\rm{\Sigma }}^2}{\rm{C}})}}{{2\underline\lambda ({C^{\top} }{\rm{\Sigma }}C)}} - 1} \right]
\end{equation}
This makes ${\delta _1} > 0 $.
Combining the inequalities (\ref{e17}) and (\ref{eq18}), we have
\[\dot V(t) \le  - \frac{{{\delta _1}}}{{{\beta _1}}}{V_1}(t) - \frac{{{\delta _2}}}{{1 + D}}{V_2}(t) \le  - \mu V(t)\]
where $ \mu  = \min \left\{ {\frac{{{\delta _1}}}{{{\beta _1}}},\frac{{{\delta _2}}}{{1 + D}}} \right\} $.

Hence, the origin of transformed system $\left( {\tilde X,\;\tilde {\cal W}} \right) $ in (\ref{e15}) is a globally and exponentially stable equilibrium point in the sense of the norm $ { {{{\| {\tilde X(t)} \|}^2} + \int_0^D {{{\tilde {\cal W}}^{\top} }(x,t)\tilde {\cal W}(x,t)dx} } } $. Finally, from the transformation (\ref{error}), we achieve exponential convergence in the sense of the norm (\ref{norm}).

From inequality (\ref{epsilon}), the maximum tolerable delay and admissible range of the observer gain in Theorem 1 are derived as
\begin{equation}\label{eq:Dmax}
\begin{aligned}
&~~~~~~~0<D<D_{max}=\frac{1}{{3\sigma }} + \sigma  - 1 \\
&0 <  \underbrace{- \frac{1}{T}{W_0}( - {\upsilon _2})}_{\varepsilon_{min}} < \varepsilon  <\underbrace{- \frac{1}{T}{W_{ - 1}}( - {\upsilon _2})}_{\varepsilon_{max}}
\end{aligned}
\end{equation}
where $$ \sigma  = {\left( {\frac{{{\upsilon _1}}}{2} + \sqrt {\frac{{\upsilon _1^2}}{4} - \frac{1}{{27}}} } \right)^{1/3}} $$ $${\upsilon _1} = \frac{{2\underline\lambda ({C^{\top} }{\rm{\Sigma }}C)(1 + 2{e^{ - 1}}) - 2{\kappa _1} \bar \lambda ({C^{\top} }{{\rm{\Sigma }}^2}C)}}{{\gamma \kappa _1^{ - 1}\bar\lambda ({C^{\top} }C)}}$$
and finally
$${\upsilon _2} = \frac{1}{2}\big( {\frac{{D(D + 1)(D + 2)\gamma \kappa _1^{ - 1}\bar\lambda ({C^{\top} }C) + 2{\kappa _1}\bar\lambda ({C^{\top} }{{\rm{\Sigma }}^2}{\rm{C}})}}{{2\underline\lambda ({C^{\top} }{\rm{\Sigma }}C)}} - 1} \big)$$
Furthermore,  $T$ is defined in {\it{Lemma}} \ref{lem1}, $ \underline\lambda $ and $ \bar\lambda $ denote the minimum and maximum eigenvalue of their corresponding matrix and the functions ${W_0}$ and ${W_{ - 1}}$ are defined in subsection \textit{A. Lambert function} in the Appendix. \carre

\subsection{Proof of {\it Theorem 2}}
 From the Lyapunov-Krasovskii functional (\ref{e16}), we have
\begin{align*}
{\varphi _1}\left( {{{\| {\tilde X(t)} \|}^2} + \int_0^D {{{\tilde {\cal W}}^{\top} }(x,t)\tilde {\cal W}(x,t)dx} } \right) \le V(t) \le {\varphi _2}\left( {{{\| {\tilde X(t)} \|}^2} + \int_0^D {{{\tilde {\cal W}}^{\top} }(x,t)\tilde {\cal W}(x,t)dx} } \right)
\end{align*}
where $ {\varphi _1} = \min \left\{ {{\beta _1},\;\rho } \right\} $ and $ {\varphi _2} = \min \left\{ {{\beta _2},\;\rho (1 + D)} \right\} $ . By virtue of (\ref{error}), we obtain
\begin{align*}
\int_0^D {{{\tilde {\cal W}}^{\top} }(x,t)\tilde {\cal W}(x,t)dx}  \le {\phi _1}\int_0^D {{{\tilde {\cal U}}^\top}(x,t)\tilde {\cal U}(x,t)dx}  + {\phi _2}{\| {\tilde X(t)} \|^2}\\
\int_0^D {{{\tilde {\cal U}}^\top}(x,t)\tilde {\cal U}(x,t)dx}  \le {\phi _3}\int_0^D {{{\tilde {\cal W}}^{\top} }(x,t)\tilde {\cal W}(x,t)dx}  + {\phi _4}{\| {\tilde X(t)} \|^2}
\end{align*}
where ${\phi _1} = 1 + {\kappa _3}$, ${\phi _2} = (1 + \kappa _3^{ - 1})\bar \lambda ({C^{\top} }C)D $, ${\phi _3} = 1 + {\kappa _4}$, ${\phi _4} = (1 + \kappa _4^{ - 1})\bar \lambda ({C^{\mathop{\rm T}\nolimits} }C)D $. Combining the above inequalities, implies
\[ {\psi _1}\left( {{{\| {\tilde X(t)} \|}^2} + \int_0^D {{{\tilde {\cal U}}^\top}(x,t)\tilde {\cal U}(x,t)dx} } \right) \le {\| {\tilde X(t)} \|^2} + \int_0^D {{{\tilde {\cal W}}^{\top} }(x,t)\tilde {\cal W}(x,t)dx}\]
\[ {\| {\tilde X(t)} \|^2} + \int_0^D {{{\tilde {\cal W}}^{\top} }(x,t)\tilde {\cal W}(x,t)dx}  \le {\psi _2}\left( {{{\| {\tilde X(t)} \|}^2} + \int_0^D {{{\tilde {\cal U}}^\top}(x,t)\tilde {\cal U}(x,t)dx} } \right)
\]
where $ {\psi _1} = \frac{1}{{\max \{ {\phi _3},\;1 + {\phi _4}\} }} $, ${\psi _2} = \max \{ {\phi _1},\;1 + {\phi _2}\} $. Finally, by solving the differential inequality $\dot V(t) \le  - \mu V(t) $ and substituting from inequality
\[{\varphi _1}{\psi _1}\left( {{{\left\| {\tilde X(t)} \right\|}^2} + \int_0^D {{{\tilde {\cal U}}^\top}(x,t)\tilde {\cal U}(x,t)dx} } \right) \le V(t) \le {\varphi _2}{\psi _2}\left( {{{\left\| {\tilde X(t)} \right\|}^2} + \int_0^D {{{\tilde {\cal U}}^\top}(x,t)\tilde {\cal U}(x,t)dx} } \right),\]
we conclude the inequality (\ref{e18}). \carre


\begin{figure}[b]
\centering
\includegraphics[scale=0.6]{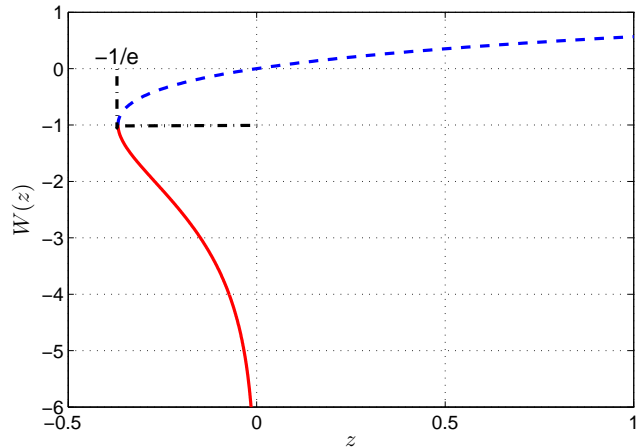}
\caption{ Solid line: ${W_{ - 1}}(z)$, Dashed line: $W_0(z)$.}
\label{fig:Lambert}
\end{figure}
\subsection{Lambert Function}
The Lambert W function is defined as the inverse of the function $y{e^y} = z$ whose solution is given by $y = W(z)$ or shortly $W(z){e^{W(z)}} = z$. For real valued $z$, if $z <  - {e^{ - 1}}$, then $W(z)$ is multivalued complex. If $- {e^{ - 1}} < z < 0$, there are two possible real values of $W(z)$: The branch satisfying $ - 1 \le W(z)$ is denoted by ${W_0}(z)$ and called the principal branch of the W function, and the other branch satisfying $W(z) \le  - 1$ is denoted by ${W_{ - 1}}(z)$ . If $z \ge 0$, there is a single real value for $W(z)$, which also belongs to the principal branch ${W_0}(z)$ \cite{c21}. The two real branches of the Lambert W function in the third-quadrant is of interest in Theorem 1. The two real branches of the Lambert W function are depicted in Figure ~\ref{fig:Lambert}.

\bibliographystyle{IEEEtran} 
\bibliography{bib}

\end{document}